Wojciech MAZURCZYK[*] and Józef LUBACZ[*]

# ANALYSIS OF A PROCEDURE FOR INSERTING STEGANOGRAPHIC DATA INTO VoIP CALLS

The paper concerns performance analysis of a steganographic method, dedicated primarily for VoIP, which was recently filed for patenting [1] under the name LACK. The performance of the method depends on the procedure of inserting covert data into the stream of audio packets. After a brief presentation of the LACK method, the paper focuses on analysis of the dependence of the insertion procedure on the probability distribution of VoIP call duration.

## 1. INTRODUCTION

Communication networks steganography is a method of hiding secret data inside usual data transmitted by users, so that the hidden data cannot be detected (in an ideal case) by scanning the data flow by a third party. A new steganographic method called LACK (Lost Audio PaCKets Steganography) was recently proposed and filed for patenting [1] in Poland. The method is described in some detail in [2].

A detailed review of steganographic methods that may be applied for IP telephony can be found in [2] and [9]. In general, the methods can be classified into the two following groups:
- Steganographic methods that modify packets: network protocol headers or payload fields. Examples of such solutions include modifications of free/redundant headers' fields for IP, UDP or RTP protocols during conversation phase and signalling messages in e.g. SIP[10]. Information hiding which is based on affecting packets' payload usually uses digital audio watermarking algorithms (e.g. DSSS [13] and QIM [14]),
- Steganographic methods that modify packets' time relations. Examples of such solution: affecting sequence order of RTP packets [11] and modifying their inter-packet delay [12].

With respect to the two above groups LACK is a hybrid steganographic method since it modifies both packets' content and their time dependencies.

In general, the LACK method is intended for a broad class of multimedia, real-time applications, but its main foreseen application (at least for now) is VoIP. The proposed method utilises the fact that for usual multimedia communication protocols like RTP (Real-Time Transport Protocol) [3] excessively delayed packets are not used for reconstruction of transmitted data at the receiver (the packets are considered useless and discarded). The main idea of LACK is as follows.

[*] Warsaw University of Technology, Faculty of Electronics and Information Technology, Institute of Telecommunications, 15/19 Nowowiejska Str. 00-665 Warsaw, Poland, {W.Mazurczyk, J.Lubacz}@tele.pw.edu.pl

At the transmitter, some selected audio packets are intentionally delayed before transmitting. If the delay of such packets at the receiver is considered excessive, the packets are discarded by a receiver not aware of the steganographic procedure. The payload of the intentionally delayed packets is used to transmit secret information to receivers aware of the procedure so no extra packets are generated. For unaware receivers the hidden data is "invisible".

The effectiveness of LACK depends on many factors such as the details of the communication procedure (in particular the type of codec used, the size of the voice frame, the size of the receiving buffer, etc.) and on the network QoS (packet delay and packet loss probability).

No real-world steganographic method is perfect – whatever the method, the hidden information can be potentially discovered. In general, the more hidden information is inserted into the data stream, the greater the chance that it will be detected, e.g. by scanning the data flow or by some other steganalysis methods. Moreover, the more audio packets are used to send covert data, the greater the deterioration of the quality of the VoIP connection. Thus the procedure of inserting hidden data should be carefully chosen and controlled in order to minimize the chance of detecting inserted data and to avoid excessive deterioration of the QoS.

To avoid excessive deterioration of the QoS lost packet ratio must be kept below certain accepted level. This level depends on the speech codec used. For example, according to [8], maximum loss tolerance is 1% for G.723.1, 2% for G.729A, 3% for G.711 codecs. If special mechanism to deal with lost packets at the receiver is utilized, like the PLC (Packet Loss Concealment) [15], acceptable level of lost packets e.g. for the G.711 codec increases from 3% to 5%. Thus this value provides us with upper limit for transmission rate.

In general, the amount of steganographic data using LACK depends on the acceptable level of packet loss. For example, for the G.711 speech codec with data rate 64 kbit/s and data frame size of 20 ms, if the packet loss probability introduced for LACK purposes is 0.5%, then under condition that packet losses do not exceed acceptable level, the theoretical hidden communication rate is 320 b/s.

In the present paper we shall focus on the hidden data insertion rate $IR$ [bits/s]. Obviously, $IR$ depends on the amount of hidden data to be sent and on the call duration. In principle, the call duration may be adjusted to the amount of hidden data to be sent. This however could cause that the distribution of calls applying LACK differs from the call duration distribution of LACK-less calls, and as a consequence make LACK vulnerable to statistical steganalysis based on call duration. Thus rather than adjusting the call duration to the amount of hidden data to be sent, it is preferable to adjust the hidden data insertion rate $IR$ to LACK-less calls duration distribution. This, in turn, requires making $IR$ dependent on that distribution. Obviously, this is not an important question in case of sporadic LACK use; it becomes important in case of a predefined group of frequent LACK users. In the present paper we focus on such a case.

Moreover, in the presented analysis we consider the dependence of $IR$ of a particular call on the elapsed time of that call, i.e. we consider $IR$ that can (potentially) be made time dependent, adjusted to the foreseen residual call duration. As shown in our analysis, such time-dependent $IR$ procedure potentially allows for decreasing the $IR$ during the call duration, compared to the $IR$ at call initiation time. In effect, the negative influence of LACK on QoS can be decreased. Such an effect can be achieved for call duration distributions with coefficient of variation greater than 1; available experimental data concerning VoIP call duration distributions seem to indicate that this is a realistic assumption for real-life VoIP calls.

It should be emphasised that the LACK procedure introduced in this paper can be utilized by decent LACK users who use their own VoIP calls to exchange covert data, but also by intruders who are able to covertly send data using third party VoIP calls (e.g. in effect of earlier successful attacks by using trojans or worms or by distribution of modified version of a popular VoIP software). This is a usual tradeoff requiring consideration in a broader steganography context which is beyond the scope of this paper.

## 2. THE VoIP CALL DURATION DISTRIBUTION

For PSTN the call duration probability distribution was well known based on extensive experimental research. For many decades an exponential distribution was assumed a good enough approximation for engineering purposes. VoIP is a relatively new service and thus only few reliable experimental data is available, so in many research papers concerning IP voice traffic (e.g. [4], [5], [6]) the exponential call duration is still assumed. Current experiments prove however that this assumption is far from being realistic.

Birke et al. [7] captured real VoIP traffic traces (about 150 000 calls) from FastWeb, an Italian telecom operator. The obtained call duration probability distribution is reproduced in Fig. 1 with a solid line. To illustrate qualitatively the degree in which the experimental results differ from the exponential distribution and some other chosen distributions (hyperexponential and log-normal) these were drawn with broken lines in Fig.1. As can be seen, the differences are considerable and no straightforward approximation of the experimental data with standard distributions is available. In particular, the exponential distribution is far from being realistic.

The experimental data yield average call duration $E(D) = 117.31$ and standard deviation $\sigma(D) = 278.74$, thus the coefficient of variation $C_v = \sigma(D)/E(D) = 2.37$ (for exponential distribution $C_v = 1$).

To achieve an analytic approximation of the experimental data a combination of some standard distributions can be used, for example:

$$f_D(x) = \begin{cases} \dfrac{1}{1.55x\sqrt{2\pi}} e^{-\dfrac{(\ln(x)-3.8)^2}{4.805}} & \text{for} \quad 0 \leq x < 27.5 \\ 0.000114 e^{-0.00114x} + 0.027252 e^{-0.03028x} & \text{for} \quad 66.5 < x \leq 27.5 \\ \dfrac{1}{1.55x\sqrt{2\pi}} e^{-\dfrac{(\ln(x)-3.8)^2}{4.805}} & \text{for} \quad 66.5 \leq x \leq 455 \end{cases} \quad (1)$$

The above analytic approximation is quite complex and of little practical use for our purposes, i.e. for establishing the dependence of the insertion rate *IR* on some simple enough characterization of the call duration distribution. Our guess was that this can be achieved through characterizing a considerably wide range of call duration distribution types with the coefficient of variation $C_v$ and then expressing the *IR* through $C_v$.

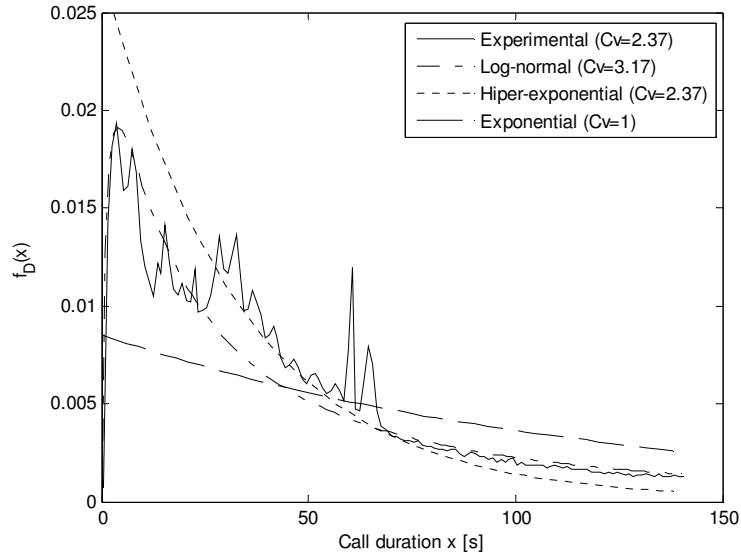

Fig. 1. VoIP call duration – comparison of experimental data with selected probability distributions

A reasonably wide range of call distribution types can be achieved and effectively analyzed with the 2-parameter Weibull distribution, with appropriately chosen parameters: the shape parameter $k > 0$ and the scale parameter $\lambda > 0$. The complementary cumulative probability distribution function $(\overline{F}_D)$ and probability density function $f_D$ are as follows:

$$\overline{F}_D(x;k,\lambda) = e^{-\left(\frac{x}{\lambda}\right)^k}$$

$$f_D(x;k,\lambda) = \frac{k}{\lambda}\left(\frac{x}{\lambda}\right)^{k-1} e^{-\left(\frac{x}{\lambda}\right)^k}$$

(2)

The $\lambda$ parameter was set so to achieve the above experimental average call duration time $E(D) = 117.31$ and the k parameter was varied so to obtain a wide range of $C_v$ values. In Table 1 the analyzed values are summarized.

| Weibull parameters | k=3.4, $\lambda$=130.57 | k=2, $\lambda$=132.37 | k=1.2, $\lambda$=124.71 | k=1, $\lambda$=117.31 | k=0.8, $\lambda$=103.54 | k=0.6, $\lambda$=77.97 | k=0.5, $\lambda$=58.65 | k=0.4, $\lambda$=35.3 |
|---|---|---|---|---|---|---|---|---|
| $C_V$ | 0.32 | 0.52 | 0.84 | 1 | 1.26 | 1.76 | 2.23 | 3.14 |

Table 1. Chosen Weibull distribution parameters k and $\lambda$ and corresponding $C_v$ values.

In Fig. 2 the Weibull probability distribution is depicted for the parameters from Tab. 1 to illustrate the resulting wide range of distribution shapes. Note by the way that for $k = 1$ the Weibull distribution equals the exponential distribution ($C_v = 1$), for $k = 2$ it becomes the Rayleigh distribution ($C_v = 0.52$) and for $k = 3.4$ it resembles the normal distribution ($C_v=0.32$).

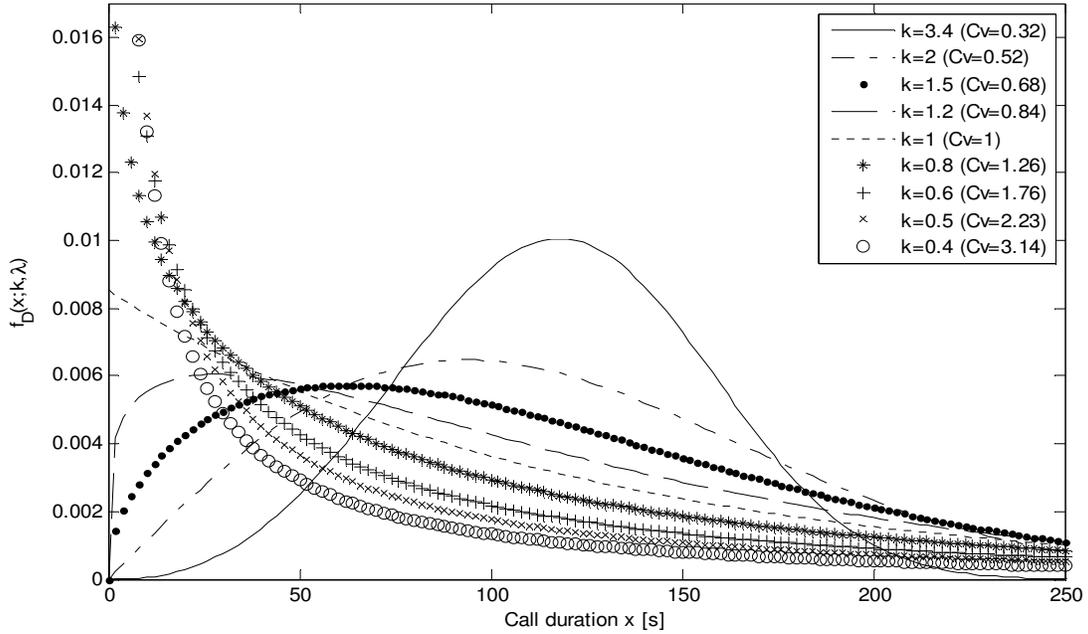

Fig. 2. Weibull distribution for various $k$, $\lambda$ and $C_V$

## 3. TIME DEPENDENT INSERTION RATE

For an arbitrary instant of a call the average residual call duration is well know to be equal

$$E(R) = \frac{E(D^2)}{2E(D)} \qquad (3)$$

or equivalently

$$E(R) = \frac{C_v^2 + 1}{2} E(D) \qquad (4)$$

Suppose that at the beginning of a call the insertion rate is set to $IR = S/E(D)$, where $S$ is amount of steganographic data to be sent covertly. As mentioned in section 1, if $C_v>1$, and thus $E(R) > E(D)$, which seems to be the case for VoIP calls as indicated above, then beginning from some arbitrary instant of the call we may decrease the insertion rate to $IR = S/E(R)$, which is beneficial from the point of view of QoS and resistance to detection of the hidden data.

The above indicates that it reasonable to make the insertion rate dependent on the elapsed time of a call. It is nevertheless not practical to use the classical definition of residual call duration since it involves an arbitrary time instant and not the current call duration. We are rather interested in the expected call duration on condition it has already lasted $t$ units of time:

$$E(D|D>t) = \frac{1}{P(D>t)} \int_t^\infty x f_D(x) dx = t + \frac{1}{\overline{F}_D(t)} \int_t^\infty \overline{F}_D(x) dx \qquad (5)$$

thus

$$t \le E(D|D>t) \le \frac{E(D)}{\overline{F}_D(t)} \qquad (6)$$

For the Weibull distributions considered in the previous section

$$E(D|D>t) = t + e^{\left(\frac{t}{\lambda}\right)^k} \int_t^\infty e^{-\left(\frac{x}{\lambda}\right)^k} dx \qquad (7)$$

and

$$t \le E(D|D>t) \le e^{\left(\frac{x}{\lambda}\right)^k} \lambda \Gamma\left(1+\frac{1}{k}\right) \qquad (8)$$

For the parameters from Tab.1 we obtain results shown in Fig.3. The figure illustrates also the *E(D|D>t)* function for the experimental data presented in the previous section.

The curves from Fig. 3 may be approximated with good accuracy as follows:

$$E(D|D>t) \approx 1.32 C_v + t\sqrt{C_v} + 0.59 \qquad (9)$$

Using this simple approximation we may establish a time-dependent insertion rate we were looking for. Suppose that the amount of remaining steganographic data to be sent at time *t* is $S_R(t)$. Then the insertion rate at time *t* may be expressed as

$$IR(t) = \frac{S_R(t)}{E(D|D>t)} \qquad (10)$$

Finally, we may modify the above *IR(t)* with a correction factor *CF<1* to reflect the fact that the LACK procedure may decrease to some extent the QoS of the speech transmission and also to take account of the required robustness to steganalysis: $IR^*(t) = CF \cdot IR(t)$. This however is a very simplified solution of the problem.

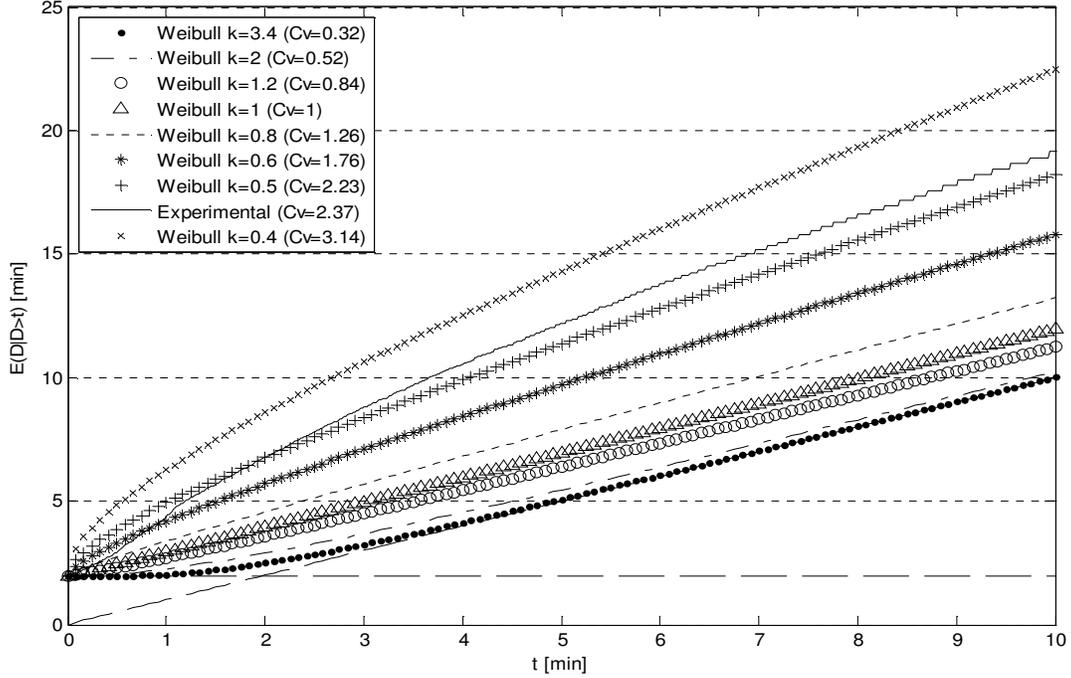

Fig. 3. *E(D|D>t)* for different Weibull distributions

Based on results presented in Fig. 3 and equation 10, for chosen Weibull distributions, the *IR(t)* values are as presented in Fig. 4:

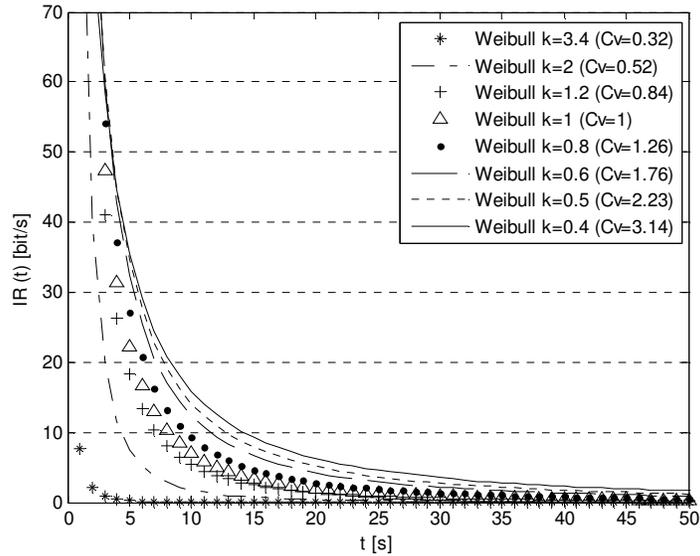

Fig. 4. *IR(t)* for different Weibull distributions for *S=1000* bits

As can be seen, call duration distributions with higher $C_v$ yield higher *IR* values. In effect distributions with higher $C_v$ allow to transmit more steganographic data.

## 4. CONCLUSIONS AND FUTURE WORK

The LACK steganographic method is a new idea which requires detailed performance evaluation. This paper is only an initial step in this direction. We have focused only on one aspect, namely the dependence of the procedure for inserting hidden data on the call duration probability distribution. It was shown that the insertion rate may be effectively made dependent on the current call duration time, and that this dependence can be expressed with good accuracy through the coefficient of variation of the call duration probability distribution. The derived formulae are simple and can be straightforwardly implemented. The effectiveness of the resulting procedure will depend on the accuracy of the estimated mean call duration and the coefficient of variation of the call duration.

The proposed procedure was made as simple as possible. A more sophisticated version of the procedure would require more detailed information about the call duration probability distribution, which might be too demanding considering the current limited experience with VoIP traffic. Nevertheless a theoretical research seems worthwhile. The authors have analyzed the problem of expressing the insertion rate function $IR(t)$ through a $P(D>x|D>t)$ distribution instead of the $E(D|D>t)$ function which was considered in the present paper. The results will be presented in a future paper.

Another task is to take into account the dependence of the $IR(t)$ function on constraints implied by QoS requirements and by the required resistance of LACK to steganalysis. In the present paper we have practically not considered this problem apart from introducing a correction factor into the $IR(t)$ which is clearly only an indication of the problem.

## ACKNOWLEDGMENTS

- This research was partially supported by the Ministry of Science and Higher Education, Poland (grant no. 3968/B/T02/2008/34).
- The authors would like to thank R. Birke, M. Mellia, M. Petracca and D. Rossi from Politecnico di Torino for sharing details of their VoIP experimental data.